\documentclass[a4paper,fleqn,usenatbib]{mnras}
\usepackage{lineno}
\usepackage{cancel}
\usepackage{ulem}

\usepackage[T1]{fontenc}
\usepackage{ae,aecompl}
\usepackage{amsmath}
\defcitealias{Markevitch_cf}{MV07}

\usepackage{graphicx}	
\usepackage{amsmath}	
\usepackage{amssymb}	

\title[$\textit{Chandra}$ observation of a cold front in Abell 2554]{$\textit{Chandra}$ observation of a cold front in Abell 2554}

\author[M. K. Erdim and M. Hudaverdi]{
M. Kiyami Erdim$^{1,2}$ and Murat Hudaverdi$^{1}$\thanks{E-mail: hudaverd@yildiz.edu.tr (M.H.)}
 \\
$^{1}$Y{\i}ld{\i}z Technical University, Faculty of Science and Art, Department of Physics, Istanbul 34220, Turkey\\
$^{2}$Y{\i}ld{\i}z Technical University, Graduate School of Natural and Applied Sciences, Istanbul 34220, Turkey}

\date{Accepted  ****. Received ****; in original form ****}

\pubyear{2018}
   \volume{**}

\begin{document}

\label{firstpage}
\pagerange{**--**}
\maketitle

\begin{abstract}
We present the evidence for the existence of substructure in the cold front cluster A2554 based on a 20.14 ks $\textit{Chandra}$ observation. 
Using centroid shift and X-ray brightness concentration parameters, we confirm that A2554 is a dynamically disturbed system.
We detect two dominant structures; a main cluster at $z=0.1108$ and a foreground northern substructure at $z=0.1082$.
The analysis reveals an X-ray surface brightness edge at $r \simeq 60$ $h^{-1}$ kpc from the cluster core. 
The thermodynamical profiles across the edge are ruling out the shock scenario.
The temperature jump (from $\sim$6 keV to $\sim$10 keV), and pressure equilibrium 
($P_{0}/P_{1} = 1.01 \pm 0.23$) across the edge, are consistent with the definition of a cold front 
with a Mach number $\mathcal{M}_1=0.94^{+0.13}_{-0.17}$.
We also observed a weak bow-shock at $\sim$100 kpc in front of the cold cloud,
corresponding an upper limit to the Mach number $\mathcal{M}_1$ $\sim$1.1. 
If the northern substructure was not related to the cold front, 
we conclude that the transonic motion of the cloud is caused by a merger, 
which was weak or occurred long ago

\end{abstract}




\section{Introduction}

Clusters of galaxies are formed by accretion and infall of smaller mass concentrations. 
The hot diffuse extended plasma, which is trapped in the gravitational potential of the cluster, is highly disturbed by these merging dynamics.  
During the merger, a significant part of the kinetic energy of substructures ($\gtrsim$10$^{64}$ ergs s$^{-1}$) 
is transformed into thermal energy of the gas and intracluster medium (ICM) therefore becomes turbulent \citep{mark99}. 

In the course of their formation, the majority of the clusters show evidences of dynamical activities at different scales. 
As a result of high-resolution spectroscopic imaging of \textit{Chandra}, 
we can study a variety of complicated transient features which was not seen before. 
One of these physical processes is called \textit{cold front}, which was first confirmed in A2142 \citep{mark2000}. 
This phenomenon was observed in many clusters subsequently by the accumulation of \textit{Chandra} data  
(e.g. A3667, \citealt{vikh2001}; RX J1720.1+2638, \citealt{moza2001}; and A1795, \citealt{mark2001}). 

The cold fronts are traced by dense, low-entropy substructures and 
generally thought to be either survived the remnants of a merger, or disturbed gas from the cluster's own cool core (\citealt{Markevitch_cf}, hereafter \citetalias{Markevitch_cf}).
They are the gas clouds moving through a hotter and less dense surrounding ICM, 
which cause contact discontinuities at the boundaries of the interaction.
The common cold fronts are found in mergers as well as in relaxed clusters around the central density peaks \citepalias{Markevitch_cf}.
Their steep density and temperature gradients show distinctive features. 
However, contrary to the cold fronts, shocks induce different thermodynamical consequences.  
The discontinuities caused by shocks are characterized by the Rankine--Hugoniot jump conditions, 
which relate the rations of the density, and the temperature before and after the shock \citep{landau}.
Across the density jump, the cold fronts show pressure equilibrium, while large pressure jumps are foreseen in shock fronts. 
For a detailed review about cold fronts and shocks in galaxy clusters the readers are referred to \citetalias{Markevitch_cf} and references therein.

\begin{figure*}
 \begin{center}
\includegraphics[width=17cm]{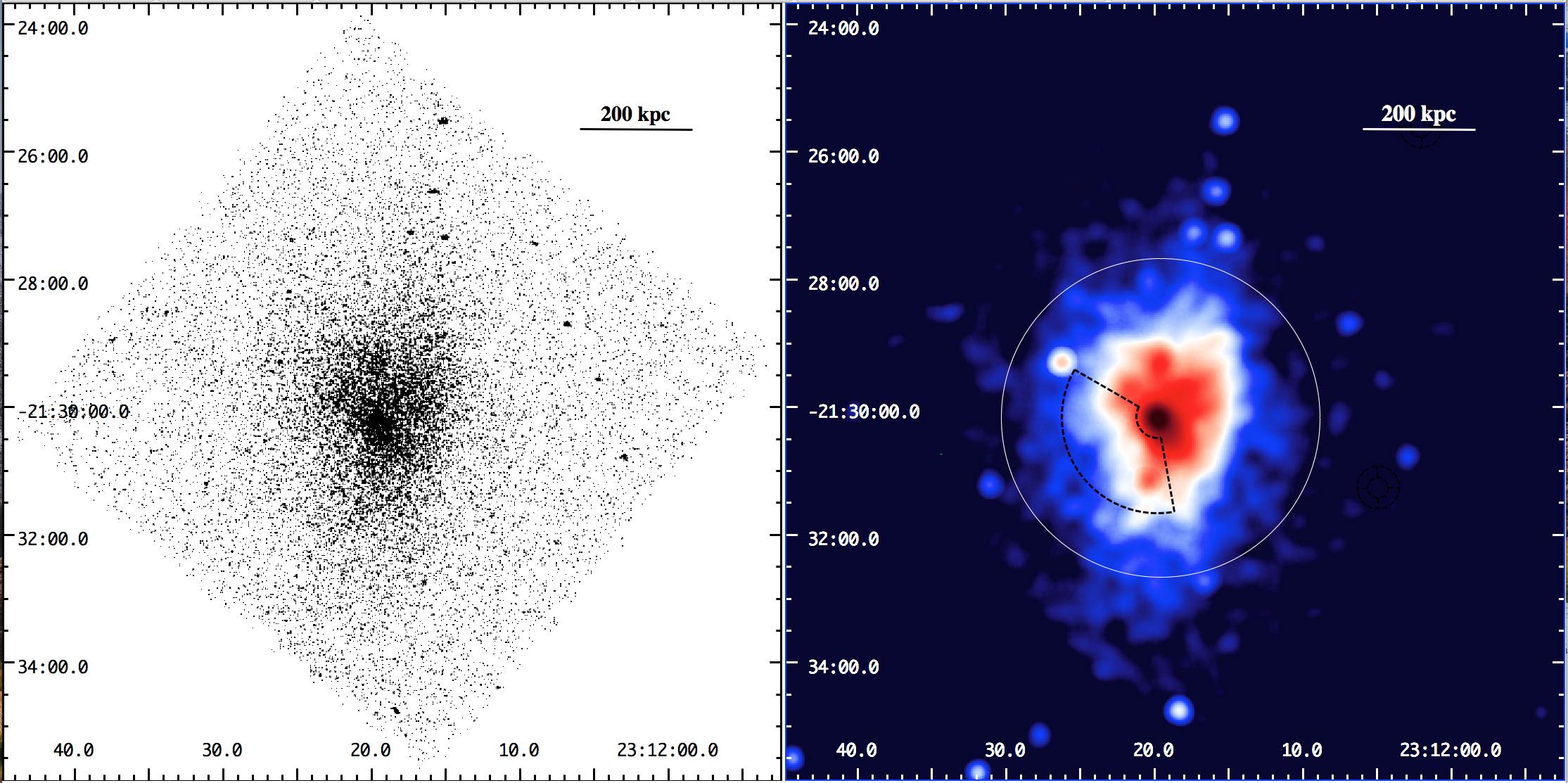}
\caption{\label{A2554_x} 
Left: $\textit{Chandra}$ ACIS-S3 $0.5-7.0$ keV energy band logarithmically spaced raw image. 
Right: adaptively smoothed, exposure--corrected and background subtracted version of the left image.
The white circle and the black sector show the regions for overall spectral analysis and the brightness discontinuity, respectively.
}
\end{center} \end{figure*}

A2554 (J2000; R.A.: $23^h 13^m 20^s.7$, decl.: $-21^d 30^m 02^s$) is a nonrelaxed galaxy cluster 
at moderate redshift $z = 0.1108$ \citep{caretta} 
with an X-ray luminosity $L_X = 1.78 \times 10^{44}$ ergs s$^{-1}$ \citep{jones}.
The cluster is one of the five richness class $R \geq 1$ clusters in a very peculiar tight knot of Aquarius supercluster \citep{batuski}. 
Based on 42 member galaxies' data, \cite{Zabludoff} determined $V_r$ = 33246$\pm$175 km s$^{-1}$ and 
velocity dispersion $\sigma_r$ = 827 km s$^{-1}$. 
The cluster has a virial mass of $0.66 \pm 0.07 \times 10^{15} M_\odot$ \citep{pear}.
Since it is on a highly dense, and crowded portion of the sky,  
the line-of-sight \textit{Einstein} IPC image looks like a trio-structure with its close neighboring clusters \citep{jones}; 
A2550 ($z = 0.1226$) 17.5 arcmin to the southwest, 
and A2556 ($z = 0.0871$) 12.5 arcmin to the southeast. 
To date, no definite bound or any filamentary structure is reported in the trio-cluster, 
but it is tempting to relate any X-ray brightness elongation to the close neighbors.  


In this work, we study the X-ray morphology and dynamical history of A2554 by analyzing a brightness edge, 
which is similar to those detected in several other clusters by Chandra (\citealt{mark2000}; \citealt{vikh2001}; \citealt{moza2001}).
This paper is organized as follows; in Section \ref{data} we describe the observation and data reduction steps.
We outline the analysis in Section \ref{analy}.
We present the morphological features, and thermodynamical properties in Section \ref{morf} and \ref{therm}, respectively.
In Section \ref{discuss}, we discuss the structure and the motion of the cloud. 
We summarize our results in Section \ref{results}.

Throughout this study, we adopt the cosmological parameters H$_{0}$=73, $\Omega_{m}$=0.27, 
$\Omega_{\Lambda}$=0.73 in a flat universe.
For this cosmology, an angular size of 1 arcsec corresponds to 1.98 kpc at the cluster redshift.
All quoted errors are derived at the 68$\%$ confidence level unless the otherwise is stated.


\section{Observation and Data Reduction}\label{data}

A2554 was observed with the Advanced CCD Imaging Spectrometer (ACIS) on board 
$\textit{Chandra}$ in a 20.14 ks observation on 2001 October 21 (Obs.ID. 1696), by using VFAINT mode. 
The focal plane temperature was set to be $-120^{\circ}$C.
The cluster center was positioned virtually on the ACIS-S3 chip with $\sim$7 arcsec offset.
Since the pointing covers almost the entire X-ray emission of the cluster in a single chip, 
in the following sections we mainly focus on the data from the S3 chip.

Using the $\textit{Chandra}$ data analysis package CIAO v4.9 software, HEAsoft v6.22.1 and CALDB v4.7.5.1, 
we followed the standard data reduction routine and removed bad pixels, and columns.  
The raw data was reprocessed by \texttt{chandra-repro} to produce level 2 event file. 
After screening procedure the total filtered effective exposure time is 19.88 ks. 
The raw uncorrected broad band $0.5-7$ keV $\textit{Chandra}$ image is presented in the left panel of Fig. \ref{A2554_x}.
The right panel of Fig. \ref{A2554_x} shows the corresponding exposure corrected, adaptively smoothed $\textit{Chandra}$ image 
highlighting the hot diffuse X-ray morphology. 
The emission appears slightly elongated in the north-south direction due to a surface brightness discontinuity edge at 
approximately 60 kpc southeast of the cluster center.
The dashed black sector indicates the region for emission discontinuity.
The large white circle shows the region where spectra were extracted for global temperature and abundance measurements,
which will be discussed in Section \ref{spect}.

\begin{figure*}
 \begin{center}
\includegraphics[width=17cm]{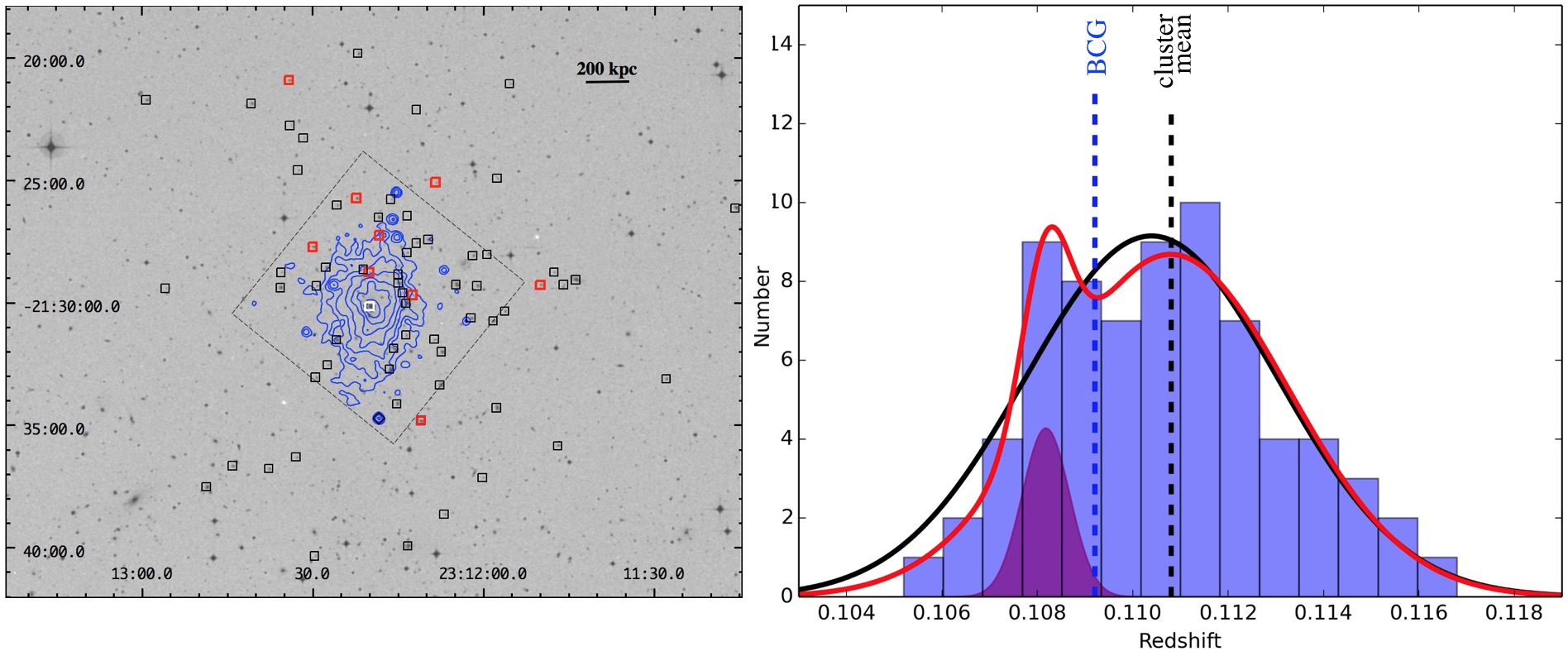}
\caption{\label{optic} Left: DSS $b$-band image centered on of A2554. 
The tilted square shows the $\textit{Chandra}$ ACIS-S3 chip field of view, 
the squares identify the galaxies that are cluster members. 
The blue logarithmically spaced contours are overlaid from the X-ray image. 
Right: Redshift distribution for 71 member galaxies.
The black curve is a single-Gaussian and
the red curve is double-Gaussian reconstruction of the fit with the potential subgroup.
}
\end{center} \end{figure*}

A Digitized Sky Survey (DSS) $b-$band image with overlaid ACIS $0.5-7$ keV energy band X-ray brightness contours 
(logarithmically spaced by a factor of 0.7) is given in Fig. \ref{optic} left panel. 
The galaxies identified as cluster members are marked with (black \& red) boxes \citep{Smith_2}.
The associated BCG is indicated by the white box, which is offset from the X-ray peak by $\sim$6 arcsec. 
ACIS-S3 field of view is indicated with the tilted square for visual aid. 
In a spectrophotometric survey of superclusters, 
the radial velocity histograms for the 48 clusters in the Aquarius region is given by \cite{Smith_2}. 
For A2554, we use the redshift data of all 71 member galaxies obtained from their 
Anglo-Australian Telescope (AAT) spectra (private communication with Smith) to identify substructures. 

The redshift histograms of relaxed clusters are known to have distributions which are close to Gaussian profiles.
Any deviation from Gaussianity is commonly interpreted as a potential substructure within the environment. 
Fig. \ref{optic} right panel shows the plot for the redshift histogram for 71 member galaxies of A2554.
The dotted lines give the cluster mean ($z=0.1108$) and the redshift of the BCG ($z=0.1092$). 
A single Gaussian profile fit (solid green line) gives the average redshift of $z=0.1104\pm0.0002$ and 
the corresponding variance $\sigma = 0.0026\pm0.0002$ with $\chi^2 = 3.56$ for 12 degrees of freedom (dof).
Nevertheless, A2554 cluster members fairly stand out, the histogram clearly diverge from a single Gaussian profile. 
Adding a second Gaussian line (solid red curve) significantly improved the fit with $\chi^2 = 1.55$ for 10 dof.
The double model reconstruction of the fit estimates a main body at $z=0.1108\pm0.0003$,
and a foreground subgroup at redshift $z=0.1082\pm0.0003$.
This redshift value for the main cluster is perfectly consistent with the previous measurements in the literature \citep{caretta,Smith_2}. 
The red boxes in Fig. \ref{optic} left panel shows 9 galaxies around
$V_r$ = 32437$\pm$90 km s$^{-1}$ interval ($0.1079<z<0.1085$). 
We notice that this redshift interval includes both the cluster and the potential subgroup members, which is difficult to distinguish.
The second Gaussian model suggests that 4 (out of 9) galaxies belong to the foreground subgroup (see Fig. \ref{optic} right panel).
It is not certain which 4 members are comprised in the subgroup,
however, considering the spatial distribution of 9 galaxies
we can safely conclude that the foreground clump is located in the north

\section{Analysis}\label{analy}
For spectral and imaging analysis, we masked any X-ray point sources which is significant at 4$\sigma$ confidence level.
Point source detection was performed using the CIAO \texttt{wavdetect} tool in the $0.5-7$ keV energy band.
After masking, the holes are replaced with a locally estimated background in the vicinity of each point sources by using the CIAO tool \texttt{dmfilth}. 
The blank sky observation, which is a stack of observations of X-ray sourceless regions in the sky is used as the background data\footnote{http://cxc.harvard.edu/contrib/maxim/acisbg/}. 
Quite often the blank sky spectrum must be normalized to the source spectrum in order to express appropriate background at the sky position. 
For this purpose, we use the $10-12$ keV energy interval in the outer part of the observation.

\subsection{Spectral Analysis}\label{spect}
The spectral analysis is performed by using pre-defined models of XSPEC v12.9.1, 
which can also be accessed from \texttt{sherpa} fitting package.
The goodness of the fit and thus the free model parameters are determined by the reduced chi-squared ($\chi^2_\nu$=$\chi^2/dof$)  test.
The global spectrum photon counts are extracted from 200 arcsec ($\sim$400 kpc) radius of the cluster center, 
which is displayed by the white circle in the right panel of Fig. \ref{A2554_x}.
The corresponding response matrix file (RMF) and energy-corrected ancillary response file (ARF) are created by \texttt{specextract} tool.
The spectra were rebinned to ensure a minimum of 20 photons bin$^{-1}$ to validate $\chi^2$ fitting.
The binned spectra were fitted in the $0.5-7$ keV range using the APEC \citep{Smith} and PHABS \citep{balu} models. 
The APEC component models a single temperature diffuse plasma, and is multiplied by PHABS, a photoelectric absorption with a revised cross-section model accounting for The Galactic Column density, which is fixed to $N_H=2.09\times10^{20}cm^{-2}$ \citep{nH}. 
The solar abundances adopted from \cite{Anders}. 
The model estimates a hot plasma with a temperature of $kT=6.56\pm{0.25}$ keV and 
an abundance of $Z=0.53\pm{0.09}$ solar with $\chi^2$ of 239.1 for 231 dof ($\chi^2_\nu=1.03$).

\section{Morphological Classification}\label{morf}

The displacement of a cluster's core is a well known, reliable indicator and has been used as 
a measure of dynamical disturbance by several authors \citep{Mohr93, poole}.
Relaxed systems expected to have concentrated distribution of hot gas with a compact core.
In the case a gravitational disturbance caused by a recent merger, the X-ray peak could deviate from the center 
and disrupted plasma would have a spread distribution. 
In order to determine A2554 morphology quantitatively, we employed two methods:
the emission centroid shift \citep{Mohr93, ohara, poole, Maughan, Mau12} and 
the brightness concentration parameter \citep{Santos}. 
\begin{enumerate}
\item The centroid shift, $w$, is a measure of standard deviation of the peak from the centroid. 
Numerical simulations \citep[e.g.,][]{poole} and systematic studies \citep[e.g.,][]{ohara}
proved that the parameter is very sensitive and a reliable indicator for cluster morphology. 
It is computed from consecutive circles centered on the X-ray peak 
by decreasing the radius of the apertures by 5\% steps from $R_{\text{ap}}$ to 0.05$R_{\text{ap}}$ as

\begin{equation}\label{eqn:cs}
 w = \Big[\frac{1}{N-1}\sum(\Delta_i - \langle\Delta\rangle)^2\Big]^{1/2} \times \frac{1}{R_{ap}},
 \end{equation} 
where $\Delta_i$ is the distance between the X-ray peak the centroid of the $i$th aperture.
\item The surface brightness concentration parameter, $c$, is introduced by \cite{Santos} and 
defined as the ratio of the peak over the ambient surface brightness, $SB$. 
The optimal $c$ ratio is found for a peak radius of 40 kpc and a cluster bulk radius of 400 kpc:

\begin{equation}\label{eqn:cp}
\centering 
 c = \frac{SB(r<40 \ kpc)}{SB(r<400 \ kpc)}.
\end{equation} 
\end{enumerate}
Adopting the equations (\ref{eqn:cs}) and (\ref{eqn:cp}) over the cluster radius, 
we calculated the centroid shift and the concentration parameter values for A2554 as 
$w=0.021\pm0.003$ and $c=0.059\pm{0.003}$, respectively.
The concentration parameter was previously determined in a systematic 41 cluster sample study 
as $c=0.066\pm{0.004}$ by \cite{Zhang}, which is consistent with our result. 

\cite{cas2010, cas2013} showed that the cluster morphological properties can be diagnosed by using the $w-c$ diagrams. 
They report a clear anti-correlation and separate radio halo (RH) clusters from clusters 
without RH with respect to the median values of $w=0.012$ and $c=0.2$.
The results also demonstrate a connection in a statistical manner between radio halo (RH) and cluster mergers for the first time:
RH clusters are dynamically disturbed systems and have $w> 0.012$ and $c< 0.2$, 
while more relaxed no RH clusters have $w< 0.012$ and $c> 0.2$ values.
A2554 possesses an extended radio source RXCJ2312.3-2130, 
which is located at 36.1 kpc from the cluster center ($z=0.110$, $F_{1.4GHz}= 37.8$ mJy; \citealt{mag}).
Considering this radio source and using the mean $w-c$ values as diagnostic reference, 
the centroid shift and the surface brightness concentration parameter calculations indicate that A2554 is not in a dynamically relaxed state.
By definition \citep{cas2010}, the cluster is classified as a morphologically disturbed merging system.  

\begin{figure}
 \begin{center}
 \includegraphics[width=8.3cm]{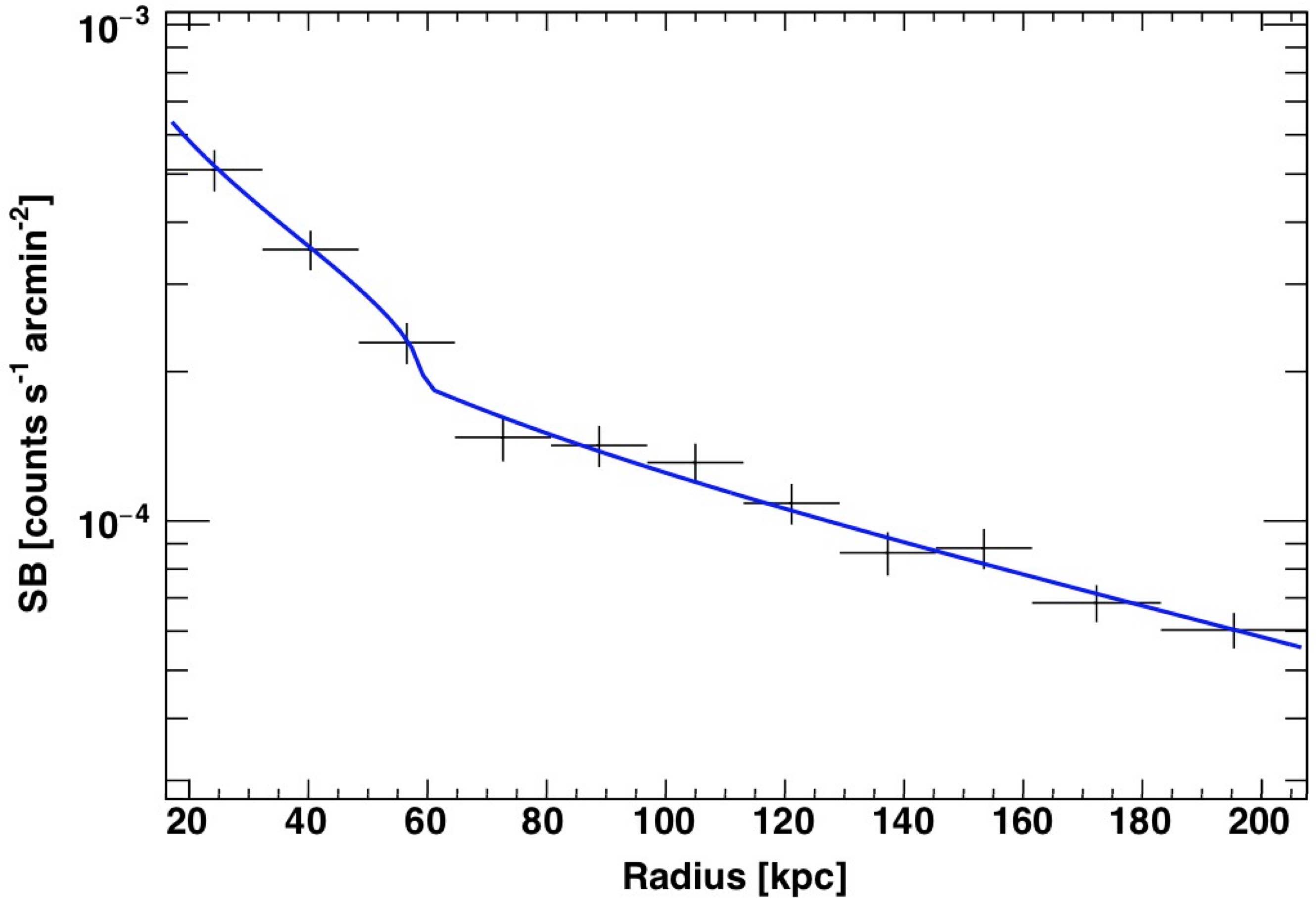}
\caption{\label{edge-plot} The surface brightness profile across the discontinuity. 
The plot is measured in southeast sector from position angle $150^{\circ}-280^{\circ}$. 
The blue curve is the broken power-law model fit.} 
\end{center}
\end{figure}

\subsection{Surface Brightness Discontinuity} \label{SB}

A2554 reveal a discontinuity in the surface brightness to the southeast from the cluster center as visually inspected in Fig. \ref{A2554_x}. 
The discontinuity is $\sim$60 kpc from the X-ray core running from position angle $150^{\circ}$ to $280^{\circ}$, 
where the position angles are measured from due west. 
In order to measure the discontinuity numerically, 
we fitted the surface brightness profile with PROFFIT v1.4 density model \citep{Eckert},
which includes a broken power-law function (\citetalias{Markevitch_cf}, \citealt{Botteon}). 
The photon counts are extracted from a sector from PA $150^{\circ}-280^{\circ}$ and 
radii 20-200 kpc across the black sector in the right panel of Fig. \ref{A2554_x}.
In the model, the downstream ($n_d$) and upstream ($n_u$) densities vary by the shock compression factor 
$\mathcal{C}\equiv{n_d}/{n_u}$ at the jump radius $r_j$:

\begin{equation}
\centering
\begin{array}{lcl}
n_d(r) = \mathcal{C}n_0(\frac{r}{r_j})^{\alpha_1} &,& (r \leq r_j) \\
n_u(r) = n_0(\frac{r}{r_j})^{\alpha_2} &,& (r > r_j), \\
\end{array}
\end{equation} 
where $n_0$ is normalization factor, $\alpha_1$ and $\alpha_2$ are the slopes of the power-laws. 

In the case of a shock, the compression factor $\mathcal{C}$ can be related to the shock Mach number

\begin{equation}
\mathcal{M}= \bigg[\frac{2\mathcal{C}}{\gamma + 1 - \mathcal{C}(\gamma -1)}\bigg]^{1/2},
\end{equation} 
by using the Rankine-Hugoniot jump conditions \citep{landau} for a thermal plasma with an adiabatic index $\gamma=5/3$,

\begin{equation}
\mathcal{C}\equiv\frac{n_d}{n_u}= \frac{4\mathcal{M}_{SB}^2}{\mathcal{M}_{SB}^2 + 3}.
\end{equation}
The corresponding surface brightness distribution with the surface brightness profile for 
the best-fitting broken power-law density model overlaid is plotted in Fig. \ref{edge-plot}. 
The surface brightness profile across this sector shows a density jump at $r_j=
59.10^{+2.85}_{-2.32}$ kpc with a compression factor of $\mathcal{C}=1.53^{+0.08}_{-0.07}$ ($\mathcal{M}_{SB}=2.4$).
In Table \ref{best-fit} we give the best fitting parameters for the broken power-law density model.

\begin{table}
\begin{center}
\caption{Fit results of the density jump.}\label{best-fit}
\begin{tabular}{cccc} 
\hline
\hline
 Model                &  $\mathcal{C}$                      & $r_j$ (kpc)     &  ${\chi}^{2}$/dof  \\
\hline
\hline
 Broken Power-law     &  1.53$^{+0.08}_{-0.07}$ & 59.10$^{+2.85}_{-2.32}$     & 	5.7/11 = 0.52\\
\hline
\end{tabular}
\label{sur-bri-log}
\end{center}
\end{table}

\begin{figure*}
 \begin{center}
 \includegraphics[width=17cm]{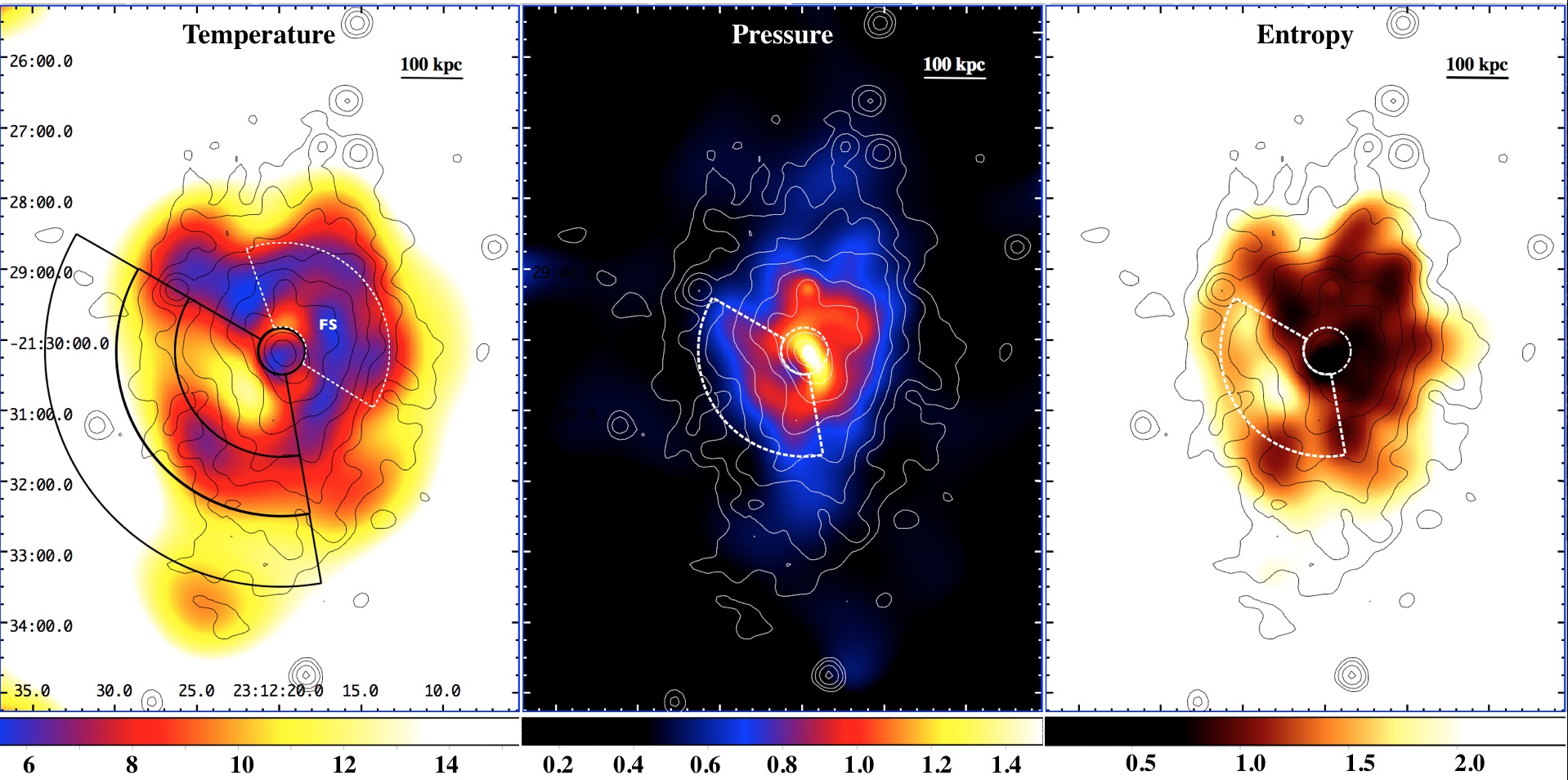}
\caption{\label{thermo} Temperature (left), Pressure (center) and entropy map (right) for A2554. 
Pseudo pressure and entropy maps are derived as $nkT$ and $kTn^{-2/3}$, 
and color-codings are estimated by $T\times(EM)^{1/2}$ and $T\times(EM)^{-1/3}$, respectively.
$0.5-7.0$ keV brightness contours are overlaid in all maps for visual aid. 
The annular sectors along the southeast direction are the regions  in which the radial profiles were extracted from.}
\end{center} 
\end{figure*}

The brightness discontinuities have been observed in several clusters of galaxies with different physical origins:
e.g. A2142 \citep{mark2000}, A3667 \citep{vikh2001}, Bullet cluster (1ES0657$-$558; \citetalias{Markevitch_cf}), A1201 \citep{owers09}.
Further insights behind the dynamics of A2554 is obtained by imaging analysis and thermodynamical properties of the ICM
to differentiate shock and cold fronts. 

\section{Thermodynamical Properties}\label{therm}
Infalling structures, merging subgroups trigger powerful kinetic energies ($10^{63}$$\sim$$10^{64}$ ergs),
which dissipate into the ICM in the form of dynamic turbulences such as shock waves and cold fronts \citepalias{Markevitch_cf}.
These disturbances change the thermodynamics of the ICM. 
In order to define the merging history of A2554 and firmly detect discontinuities in the ICM, 
we have generated thermodynamical maps (from left to right: temperature, pressure, and entropy) as shown in Fig. \ref{thermo}.
The $0.5-7.0$ keV brightness contours are overlaid on each map for visual aid.
The black and white sectors indicate the location and the direction of the surface brightness discontinuity 
from the position angle $150^{\circ}-280^{\circ}$, which was discussed in Section \ref{SB}.
The panels of Fig. \ref{thermo} shows the radial temperature, pressure, and entropy profiles,
which were extracted from the annular sectors.
The regions also highlight the part of the thermodynamical maps, where we focus our attention. 

\subsection{\label{temp} Temperature Map}
The temperature map was created with hardness ratio approximation which is a useful and powerful indicator of 
spatial variation of plasma temperature. 
The images have been extracted carefully by avoiding QDP instrumental lines, galactic
absorption at soft bands (< 0.5 keV) and point source emissions at hard bands (> 7 keV). 
To do this, we extracted two images of $0.5-1.4$ keV and $1.4-7$ keV energy (or PI) bands.
Point sources are excluded and the masked holes were filled using the surface brightness of the surrounding pixels and the CIAO tool \texttt{dmfilt}.
The soft and hard energy bands were selected so that the total count in each energy band is almost equal in order to minimize statistical errors. The corresponding backgrounds were subtracted. 
The hard image is divided by the corresponding soft image to obtain the hardness ratio map.
The ratio values in each pixel are converted to temperature (keV) values by multiplying the theoretical conversion factors. 
A similar technique for creating temperature maps from harness ratio images is described and successfully applied for 
both $\textit{XMM-Newton}$ and $\textit{Chandra}$ data (see e.g. A1758, \citealt{david}, A521 \citealt{Ferrari06}).
The conversion factors for A2554 are determined by an absorbed single thermal collisional plasma model 
$\textit{phabs} \ast \textit{apec}$ with a hydrogen absorption column density fixed at the Galactic value 
($N_H=2.09\times10^{20}$ $cm^{-2}$), and the redshift at $z=0.1108$.
The confidence of the produced map is tested by comparing spectral $\textit{kT}$ values ($T_\mathrm{spec}$) 
with the map values ($T_\mathrm{map}$) as outlined in detail by \cite{caglar17}. 
The pixel values are in good agreement and the consistency is confirmed. 

\begin{figure}
 \begin{center}
 \includegraphics[width=8.3cm]{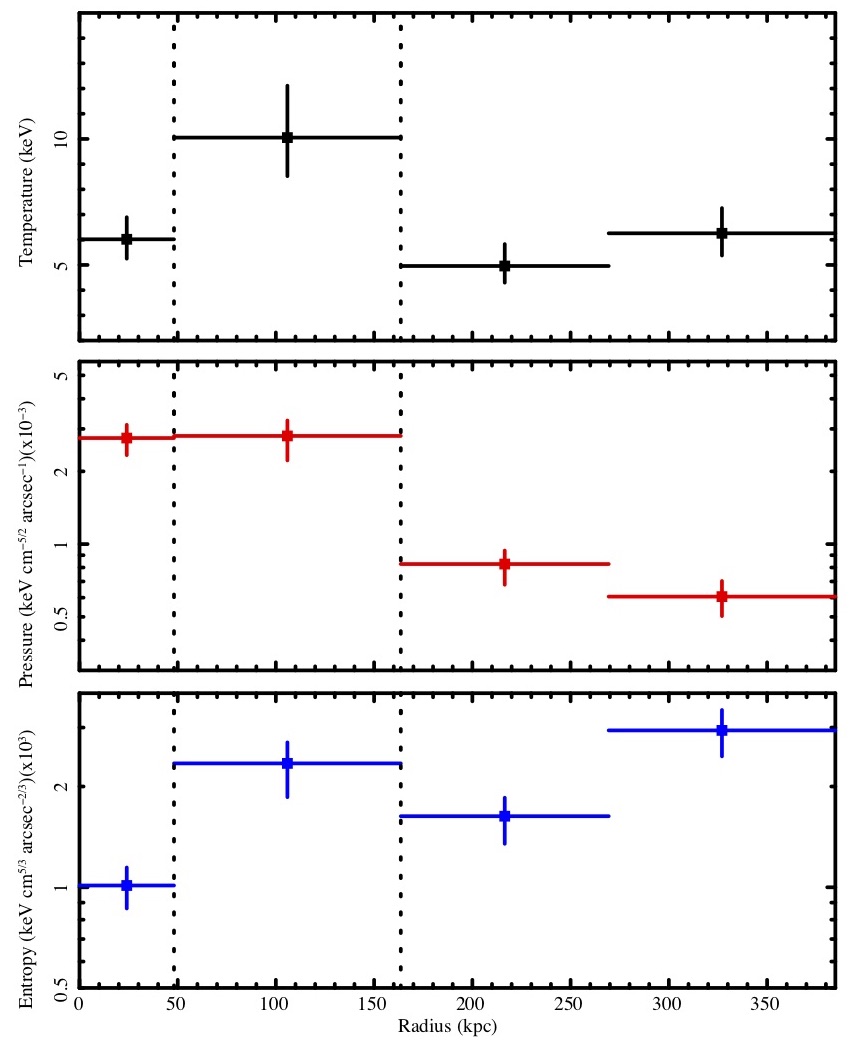}
\caption{\label{thermoprofile} Radial temperature (upper panel), pressure (middle panel) and entropy (lower panel)
distribution for the annular sector in Fig. \ref{thermo}.
The dashed vertical line indicates the positions of the discontinuities.}
\end{center} \end{figure}

The thermodynamical maps should be investigated cautiously 
since these quantities projected along the line of sight \citep[e.g.][]{mazzotta}.
To minimize projection effects it would be ideal to measure the deprojected profiles, 
however, the limited number of counts does not allow this. 
The color differences in the map represent regions which are at significantly different temperatures.
Fig. \ref{thermo} left-panel shows an evident temperature jump across the sector. 
Fig. \ref{thermoprofile} shows the profiles of temperature ($black$), pressure ($red$), and entropy ($blue$)   
across the edge along the annular sectors.

A significant temperature jump is evident across the edge (top panel Fig. \ref{thermoprofile}). 
The temperature significantly increase from $T_\mathrm{in}=6.02^{+0.87}_{-0.78}$ keV to $T_\mathrm{out}=10.05^{+2.06}_{-1.53}$ keV
along the southeast sector from PA $150^{\circ}-280^{\circ}$ centered at the X-ray peak. 
For a shock discontinuity,  the gas density jump, $r \equiv \rho_1/\rho_0$, and the temperature jump, $t \equiv T_1/T_0$, 
can be related by applying Rankine-Hugoniot jump conditions \citep{landau}. 
Here the indices 0 and 1 stand for quantities before and after the shock.

\begin{equation}
t= \frac{\zeta - r^{-1}}{\zeta - r}  
\end{equation} 
where $\zeta \equiv (\gamma + 1)/(\gamma - 1)$ with an adiabatic index $\gamma=5/3$ for a thermal plasma.
For the observed density jump at the edge, $r \sim$ 1.7 in A2554.
Assuming that the post-shock temperature $T_1\simeq$ 6 keV observed inside the edge, 
one would expect to find a $T_0 \simeq$ 4 keV gas in front of the shock.  
In fact, we observe that the temperature of the less dense gas outside the edge is 
higher ($\simeq 10$ keV) than that of the inside ($\sim 6$ keV). 
A similar case is observed for the edges in A2142  \citep{mark2000}, and A3667 \citep{briel}, ruling out the shock scenario.
The brightness edge in A2554 shows the characteristic temperature jump feature of cold fronts. 
 
\subsection{\label{PS} Pseudo Pressure \& Entropy Maps}
Since the brightness edge is not due to the shock front, we investigate pressure and entropy profiles 
in order to understand the physical reason behind the sharp edge. 
By simply using the brightness image and temperature map we can calculate pseudo pressure and entropy values
(see e.g., A2052: \citealt{plaa10}, A3667: \citealt{datta}).

The normalization value in a thermal plasma model \citep[e.g.][]{sarazin}: 

\begin{equation}\label{normalization}
\centering
\mathcal{N} = \frac{10^{-14}}{4 \pi [D_A (1+z)]^2} \int n_e n_H dV,
\end{equation} 
where $D_A$ is the angular diameter distance to the source in units of $cm$, 
$n_e$ and $n_H$ are the electron and hydrogen densities in units of $cm^{-3}$, respectively.
By dividing the normalization by the area of each pixel $EM\mathcal{=N/A}$, one can obtain the projected emission measure
which is proportional to the square of the electron density ($n_e^2$) integrated along the line of sight. 
Combining the surface brightness image and temperature map, we derived pseudo-pressure $P$ and pseudo-entropy $S$ as:

\begin{equation}
\centering
\begin{array}{lcl}
P = nkT 		& \rightarrow 	& P \sim T\times(EM)^{1/2} \\
S = kTn^{-2/3} 	& \rightarrow 	& S \sim T\times(EM)^{-1/3} \\
\end{array}
\end{equation} 
The obtained maps are shown in Fig. \ref{thermo}. 
The color-coding for pressure map ($center$), and entropy map ($right$) are arbitrary, hence the prefix pseudo is added.

In a recent work on 15 clusters observed with Chandra, \cite{Botteon}
computed pressure ratios across the detected edges and verified 
the absence of a pressure equilibrium in the shocks, 
and the presence of a pressure continuity in the cold-fronts, supporting earlier claims. 
The lack of sufficient photon counts does not allow us to study deprojected profiles for A2554.
However, the studies have also shown that projection effects along the line of sight do not have
a strong impact on the profiles \citep[e.g.][]{dasadia}. 
The pressure values before and after the edge are estimated to be 
$2.7\pm0.4$ $\times10^{-3}$ keV cm$^{-5/2}$arcsec$^{-1}$ and 
$2.8\pm0.5$ $\times10^{-3}$ keV cm$^{-5/2}$arcsec$^{-1}$, respectively. 
The gas pressure profile across the edge is clearly in equilibrium,
which leads to a pressure ratio $P_{0}/P_{1} = 1.01 \pm 0.23$ (the middle panel of Fig. \ref{thermoprofile}).
The shock fronts increase the entropy after they have passed. 
The contrary is observed for a cold front (e.g. A2052, \citealt{plaa10}).
The entropy is significantly lower in the inner side of the surface brightness discontinuity for A2554 
(the bottom panel of Fig. \ref{thermoprofile}).


\section{Discussion}\label{discuss}
We have analyzed the internal dynamics of A2554 for the fist time.
We find that the cluster is a dynamically active system and possess a surface brightness 
discontinuity at $\simeq 60$ $h^{-1}$ kpc to the southeast direction from the cluster core. 
The thermodynamical 
properties across the discontinuity rule out a shock scenario.
The temperature jump and pressure equilibrium are consistent with a "cold front". 
The cold front display the boundary between the cool cloud and the ICM.
Across all other azimuth the ICM temperature is nearly constant around 6 keV, 
the second sector has elevated temperature ($\sim$10 keV), which is formed by the motion of the cool core.
Similar radial profiles and the cold fronts are reported for RXJ0751.3+5012 \citep{rus14} and RXJ0334.2$-$0111 \citep{das2016}.

\begin{figure}
 \begin{center}
 \includegraphics[width=8.3cm]{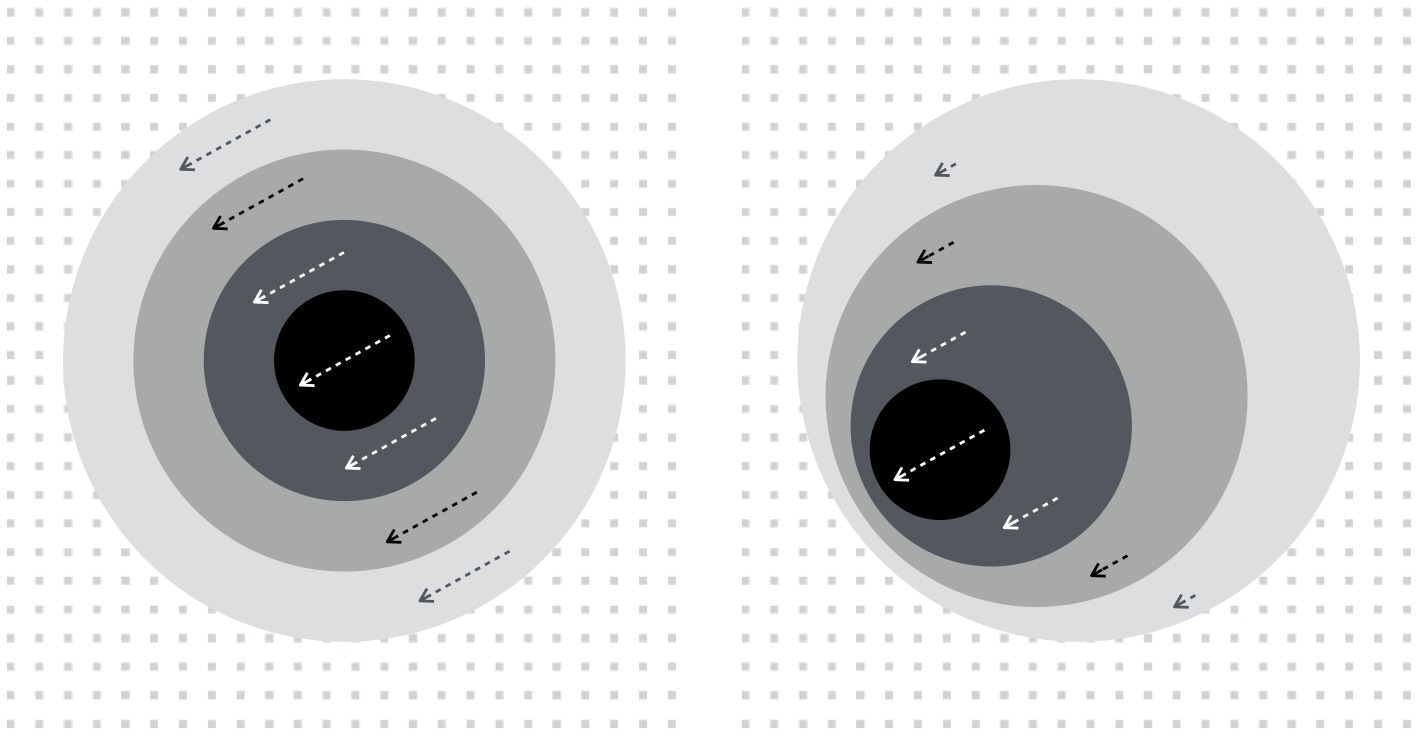}
\caption{\label{edge} The formation of a discontinuity from an initially continuous gas cloud.
The denser the gas, the less drag force it will feel,
which eventually brings the dense gas forward. }
\end{center} \end{figure}

\subsection{\label{dics1} Velocity of the cool gas}
Under normal circumstances one expect a cluster to be in hydrostatic equilibrium in its own gravitational potential. 
The gases at different radial range appear nearly at rest and in equilibrium for relaxed systems.
With this assumption, we can estimate the velocity of the front by using the deviations from spherical symmetry, and
the perturbation initiated by the motion of the cold gas. 
As the dense gas cloud moves with respect to the hot ambient gas, it feels a ram-pressure. 
The more dense the gas cloud is, the smaller the drag force and the deceleration it will experience. 
The densest gas travel faster and eventually forms a density discontinuity (\citetalias{Markevitch_cf}). 
Therefore, the hot gas in the stagnation region has a low velocity with respect to the cold front. 
A simple illustration of a formation of a discontinuity from a continuous distribution is given in Fig. \ref{edge}. 
Cosmological simulations also suggest that gravitational disturbances of a merger cause core oscillations and 
can form a cold front around the cluster core \citep{titt}.

\begin{figure}
 \begin{center}
 \includegraphics[width=8.3cm]{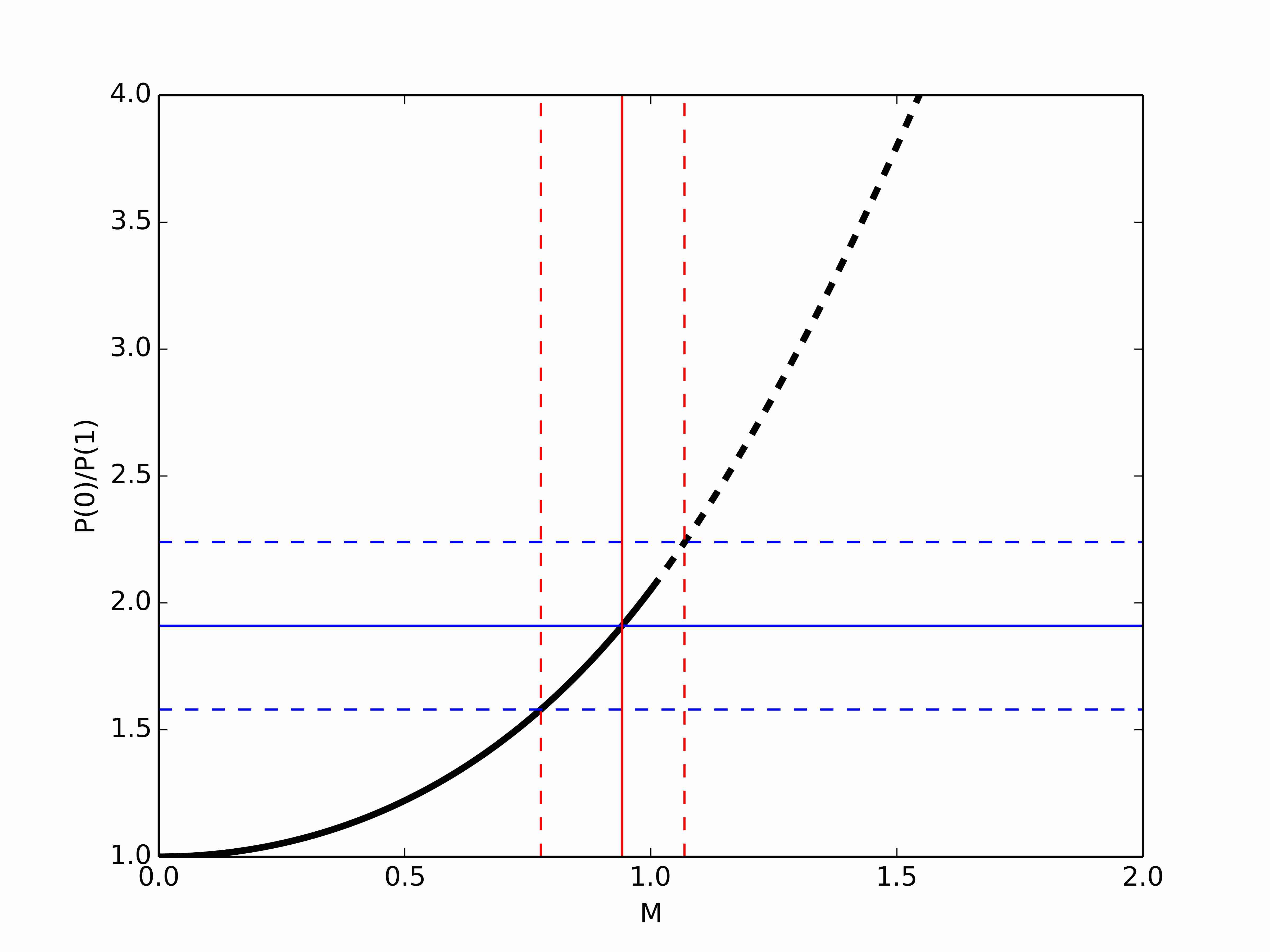}
\caption{\label{mach-edge} Ratio of pressures at the stagnation point and in the free stream, 
as a function of Mach number in the free stream (Equation \ref{land}). 
The sub- and supersonic regimes are shown as solid and dashed curved lines, respectively. 
The vertical and horizontal lines show the confidence interval for $p_{0}/p_{1}$ and $\mathcal{M}_1$ in A2554.}
\end{center} \end{figure}

The measured thermal pressure difference across the front gives the ram pressure 
which can be used to determine the front velocity with respect to surrounding gas. 
This method has been successfully applied to the cold fronts in several clusters (e.g. A3667, \citealt{vikh2001}; 
A3558, \citealt{rossetti}; A1795, \citealt{ehlert}; and A3376, \citealt{urdam}). 
The expression for the ratios of thermal pressures at the stagnation point, $P_{0}$
and in the free stream, $P_{1}$ as a function of the Mach number is defined as \citep{landau}:

\begin{equation}\label{land}
    \frac {P_{0}}{P_{1}} =\begin{cases}
    \Big[ 1 + \frac {1}{2} (\gamma -1)  \mathcal{M}_1^2   \Big]^{ \frac {\gamma}{(\gamma - 1)}}   			& , \text{$\mathcal{M}_1 \leq 1$}.\\
   \Big(\frac {\gamma + 1}{2} \Big)^{ \frac {(\gamma+1)}{(\gamma - 1)}}
    \mathcal{M}_1^2 \Big[\gamma -  \frac {\gamma-1}{2\mathcal{M}_1^2}   \Big]^{ -\frac {1}{(\gamma - 1)}} 	& , \text{$\mathcal{M}_1 > 1$}.
  \end{cases}
\end{equation}
where $\mathcal{M}_1=v/c$, $v$ is the propagation velocity in the free stream region 
and $\gamma=5/3$ is the specific-heat ratio (or adiabatic index) of the gas.
Fig. \ref{edge} shows the plot of Equation \ref{land}, which is the ratio of pressures as a function of Mach number. 
The sub- and supersonic regimes are shown as solid and dashed curved lines, respectively. 
In the equation, $P_{1}$ should be the pressure of the free stream which is uncompressed, 
unheated and undisturbed by the motion of the cold cloud. 
For this reason, we have used the region outside the edge, where there is no perturbation. 
The choice of the free stream region for A2554 is shown in the temperature map (see Fig. \ref{thermo}) with dashed-white sector labeled as FS. The spectral fits estimate
$kT_0=6.02^{+0.87}_{-0.78}$ keV and $kT_1=5.62^{+0.66}_{-0.52}$ keV, 
and thus a temperature ratio $T_{0}/T_{1} = 1.07\pm0.18$.
The normalization values of the thermal models give the emission measure for each region 
which is proportional to the density square (see Equation \ref{normalization}).
The best fit parameters of the spectral fitting yield a density ratio $n_{0}/n_{1} = 1.79\pm0.08$.
The pressure at the stagnation point, and at the free stream are calculated to be 
$p_{0}=0.22\pm0.03$ ($\times10^{-3}$) keV cm$^{-3}$ and 
$p_{1}=0.11\pm0.01$ ($\times10^{-3}$) keV cm$^{-3}$, respectively. 
The observed ratios of density and temperature imply a pressure ratio of $p_{0}/p_{1}=1.91\pm0.33$, 
which corresponds to a transonic motion, with a Mach number $\mathcal{M}_1=0.94^{+0.13}_{-0.17}$.
The cold gas cloud in A2554 is moving with a transonic velocity of $v=1151^{+173}_{-208}$ km s$^{-1}$ relative to the external gas. 
In Fig. \ref{mach-edge}, the measured pressure ratio $p_{0}/p_{1}$ and 
the corresponding Mach number $\mathcal{M}_1$ value in A2554 are shown by the horizontal and the vertical lines, respectively. 
The dotted lines show the confidence intervals. 

\subsection{Possible bow-shock}
If the cold front velocity were supersonic, a bow shock is expected at some distance in front of the moving cold cloud.
The compressed and shock-heated ICM forms a second discontinuity in the surface brightness and the temperature along the motion. 
We have searched for the possible bow-shock evidences formed by the transonic cold front in A2554. 
The surface brightness profile is extracted along the same annuli (the right panel of Fig.\ref{A2554_x}). 
A significant brightness discontinuity was not, however, observed up to 250 kpc in A2554.  
The slope in the temperature profile, on the other hand, changes from 
$T_{1}=4.95^{+0.87}_{-0.66}$ keV to $T_{2}=10.05^{+2.06}_{-1.53}$ keV, 
in which the subscripts 1 and 2 refer to the preshock and postshock gas.
In the case for A2554, the shock distance (160 kpc) from the leading edge of the body ($\sim$60 kpc) is about 100 kpc. 
The relation between the velocity of the cold front and the distance of the potential bow-shock formed by this propagation is 
estimated by using the method of Moeckel (1949)\footnote{http://naca.central.cranfield.ac.uk/reports/1949/naca-tn-1921.pdf}.
The assumptions of the method on the geometry and the symmetry of the system introduces uncertainties, 
but we can use the method as described in \cite{vikh2001} to determine an upper limit to the Mach number of the cold cloud.
The observed distance of 100 kpc corresponds to a Mach number $\mathcal{M}_1$ $\approx$ 1.1, 
in good agreement with the low Mach number estimated from the pressure ratio in \S \ref{dics1}. 
Therefore, we can conclude that the motion of the cloud is transonic, with a Mach number in the range 0.94$-$1.1.

We note that the motion of the cold front is slightly supersonic, which may not form a strong bow-shock.
The projection effects offer another reasonable explanation for the insignificancy of the secondary brightness jump. 
 If the propagation of the cloud is not perpendicular to the line of sight, 
 the large angles from the sky plane would also smear the observed bow-shock.

\section{results \& summary}\label{results}
We have presented the analysis results of an archival Chandra observation of the unrelaxed cluster A2254, 
which include thermodynamical maps and radial profiles across a remarkable southeast surface brightness edge.
Our main results are: 
\begin{enumerate}

\item The X-ray image is slightly elongated in the north-south direction perhaps 
showing it must be in the throes of a merger (Fig. \ref{A2554_x}).

\item Based on the optical spectroscopic redshifts data of 71 member galaxies from their AAT spectra, 
A2554 redshift histogram shows a clear deviation from Gaussianity. 
Two Gaussian models predict a (northern) foreground substructure with $z = 0.1082\pm0.0003$, 
and the main body at $z = 0.1103\pm0.0003$ (see Fig. \ref{optic}).

\item 4 (out of 9) galaxies in $0.1079<z<0.1085$ redshift interval belong to a northern subgroup (see Fig. \ref{optic}).
The slight asymmetry in the X-ray brightness image (Fig. \ref{A2554_x}) is rather small for major dynamical processes, 
and thus this is probably the first encounter of the substructures.
This does not constitute a strong proof, but the physical origin for the brightness discontinuity can be related to this foreground substructure. 

\item The centroid shift, and the surface brightness concentration parameters are 
measured to be $w = 0.021\pm0.003$, and $c = 0.059\pm0.003$, respectively. 
Using the $w-c$ diagrams with respect to the median values of $w=0.012$ and $c=0.2$ as morphological diagnostics \citep{cas2010}, 
A2554 is classified as dynamically active system. 

\item We fitted a broken power-law model to describe the surface brightness profile. 
The surface brightness discontinuity is defined at $r\sim60$ kpc to the southeast direction 
with a compression factor of $\mathcal{C}=1.53^{+0.08}_{-0.07}$ (Fig. \ref{edge}). 
The detected sharp discontinuity is an indicator to probe the presence of a dynamical instability (shock or cold front),  
which is generated by a merger, and thus A2554 is definitely not a relaxed system.  

\item The temperature, pressure and entropy profiles across the discontinuity are ruling out shock front scenario. 
The temperature jump (from $6.02^{+0.87}_{-0.78}$ keV to $10.05^{+2.06}_{-1.53}$ keV), 
the pressure equilibrium ($P_{in}/P_{out} = 1.01 \pm 0.23$), 
and the lower entropy in the inner side of the edge are consistent with the definition of a cold front. 

\item The gas pressure at the stagnation point is $1.91 \pm 0.33$ times higher than that at the free stream. 
The amplitude of this pressure jump requires a transonic motion of the cool gas 
with the Mach number $\mathcal{M}_1=0.94^{+0.13}_{-0.17}$.

\item The slight sonic motion of the front is confirmed by a potential bow-shock about 100 kpc from the cold front.
The observed distance corresponds to an upper limit $\mathcal{M}_1$ $\approx$ 1.1 of the cloud. 

\end{enumerate}

The analysis presented here supports the view that A2554 is definitely not a relaxed system, 
and the brightness edge in southeast is due to a transonic cold front. 
Our optical and X-ray analysis indicate that northern sub-clump is currently heading toward the main cluster core.
Disturbance caused by this aggregation is observed as the motion of the displaced gas from the cluster's own cool core. 
If the northern sub-structure was not physically related to the edge, 
the real cause of distribution was either weak or occurred long ago.


\section*{Acknowledgement}

The authors thank the anonymous referee for constructive comments and suggestions.
We thank R. E. Smith for kindly providing us with the detail information of 71 member galaxies in A2554 that are used in this paper. 
We would like to acknowledge financial support from the Scientific and 
Technological Research Council of Turkey (T\"{U}B\.{I}TAK) project number 113F117. 
The work is also supported by YTU Scientific Research \& Project Office (BAP) 
funding with contract numbers 2013-01-01-YL01 and 2013-01-01-KAP04. 


\bsp	
\label{lastpage}
\end{document}